\documentclass[aps,prl,twocolumn,longbibliography]{revtex4-1}
\usepackage{amsmath}
\usepackage{amsfonts}
\usepackage{amssymb}
\usepackage{graphicx}
\usepackage{float}
\usepackage[dvipsnames]{xcolor}
\usepackage{tabularx,booktabs}
\usepackage{lipsum}

\usepackage{todonotes}
\usepackage{soul}

\begin{document}

\title{Skyrmion Spin Ice in Liquid Crystals}
\author{Ayhan Duzgun and Cristiano Nisoli}
\affiliation{Theoretical Division, Los Alamos National Laboratory, Los Alamos, NM, 87545, USA}

\begin{abstract}
We propose  the first  Skyrmion Spin Ice, realized via confined, interacting liquid crystal skyrmions. Skyrmions in a chiral nematic liquid crystal  behave as quasi-particles that can be dynamically confined, bound, and created or annihilated individually with ease and precision. We show that these quasi-particles can be employed to realize binary variables that interact to form ice-rule states. Because of their unique versatility, liquid crystal skyrmions  can open entirely novel avenues in the field of frustrated systems. More broadly, our findings also demonstrate the viability of LC skyrmions as elementary degrees of freedom in the design of collective complex behaviors.

\end{abstract}

\maketitle 

Artificial spin ices (ASI)~\cite{Wang2006,tanaka2006magnetic,Libal2006,ortiz2016engineering,libal2009,Latimer2013,Trastoy2014freezing,Nisoli2013colloquium,heyderman2013artificial,nisoli2018frustration} are frustrated materials modelled as arrays of interacting, frustrated, binary variables arranged along the edges of a lattice. At the vertices, where these Ising spins meet, their configurations  obey  the ice rule~\cite{bernal1933theory,Pauling1935}, which often leads to various forms of {\it constrained disorder}. ASIs  can be designed for a wide variety of unusual emergent behaviors~\cite{nisoli2018frustration} often not found in natural materials~\cite{gilbert2016frustration,nisoli2017deliberate}. 
Their seminal~\cite{Wang2006,tanaka2006magnetic} and to this day most explored~\cite{Nisoli2013colloquium,heyderman2013artificial,nisoli2018frustration} realizations employ lithographically fabricated, magnetic nanoislands. Nonetheless, the same set of ideas behind these materials extend beyond magnetism, and spin ice physics has been exported to other platforms, such as superconductors~\cite{libal2009,Latimer2013,Trastoy2014freezing,wang2018switchable}, confined colloids~\cite{ortiz2016engineering,libal2018ice,ortiz2019colloquium}, magnetic skyrmions~\cite{ma2016emergent}, and elastic metamaterials~\cite{meeussen2019topological}.

In this work we demonstrate numerically Liquid Crystals (LC) as a new, timely platform for spin ice physics~\cite{duzgun2019artificial}. By confining liquid crystal skyrmions in binary traps with two preferential positions at the ends~\cite{Libal2006} we  recreate  Ising spin variables. Then their frustrated mutual repulsion leads to the ice rule~\cite{Libal2006,nisoli2018unexpected}.

Nematic LC are typically made of elongated molecules which can access phases of orientational order but no spatial order. They  exhibit a random distribution of their centers of masses yet with the alignment of their principal axis along a local director $\hat n(\vec{x})$. Their nematicity can be captured by a traceless tensor $Q_{\alpha,\beta}=S\left(3\hat n_{\alpha} \hat n_{\beta}-\delta_{\alpha,\beta}\right)/2$, $S$  being the so-called {\it scalar order parameter} quantifying  orientational order.

\begin{figure}[t!]
\includegraphics[width=\columnwidth]{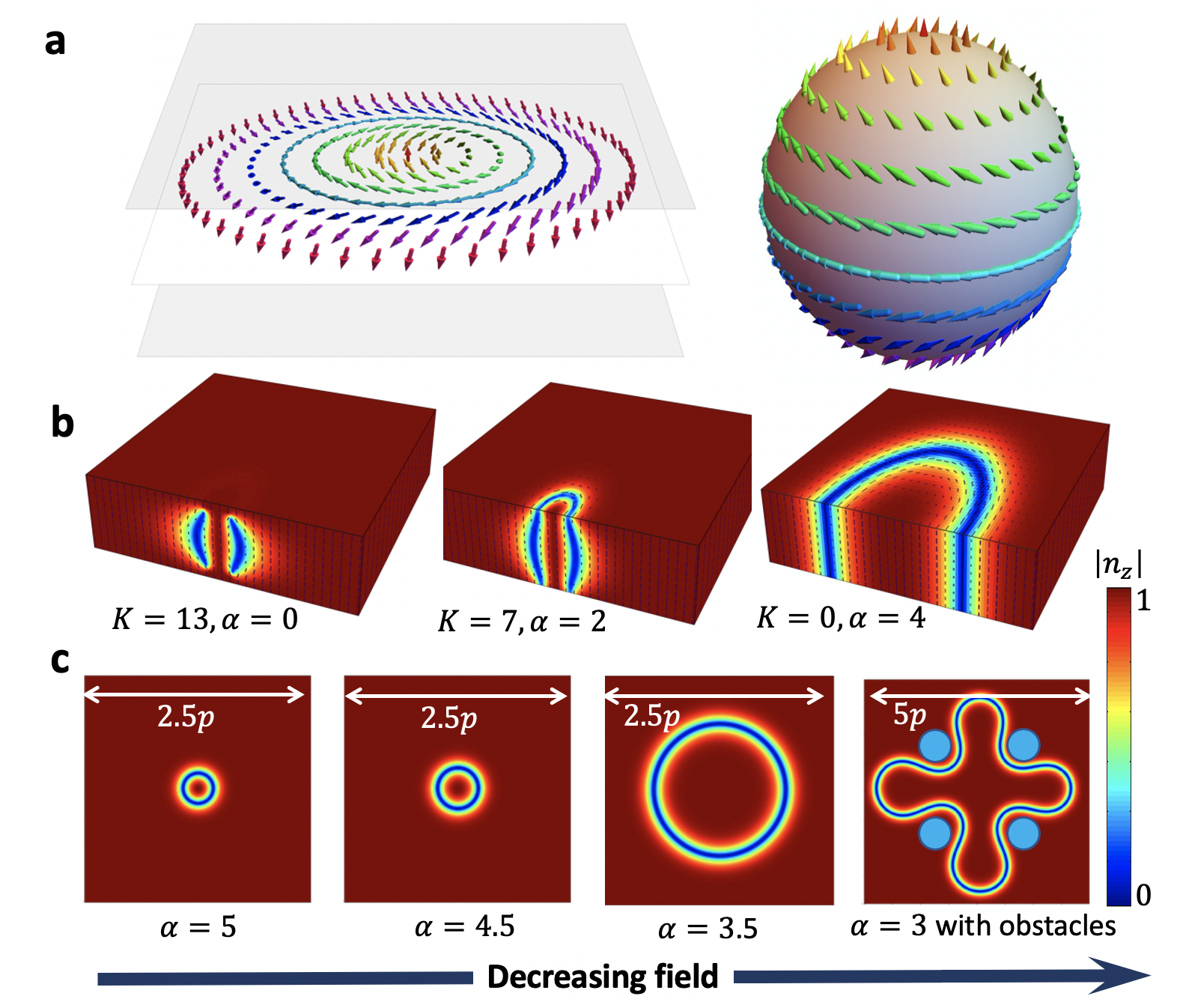}
\caption{{{\bf Skyrmion structure.}} \textbf{(a)} Mid-layer of a vectorized skyrmion in LC cell with homeotropic anchoring and mapping of directors on to the surface of a sphere \textbf{(b)} External electric field (coupling coefficient $\alpha$) and surface anchoring (coupling coefficient $K$) can be used in various proportions to sustain long lived skyrmions. Different skyrmion shapes appear depending on $K$ and $\alpha$. The ratio of the cell thickness to the cholesteric pitch is $N_z/p\approx 0.36$. \textbf{(c)} Top view of 2D skyrmions ($K=0$) stabilized by background field (stability range $5.5>\alpha >3$ as shown in \cite{duzgun_SkPhase} ). Reducing field strength results in increased skyrmion size and increased deformability near obstacles produced by strong vertical alignment (blue circles). }
\label{fig1}
\end{figure}

Our LC cell (Fig.~\ref{fig1}a)  consists of a \emph{chiral nematic} LC confined between two parallel surfaces. The  system is successfully described via a phenomenological free energy {per unit volume}
\begin{align}
f&=\frac{1}{2}a \text{Tr}\left(Q^2\right)+\frac{1}{3}b \text{Tr}\left(Q^3\right)+\frac{1}{4}c\left[\text{Tr}\left(Q^2\right)\right]^2
\nonumber\\
&+\frac{1}{2}L\left(\partial _{\gamma }Q_{\alpha \beta }\right)\left(\partial _{\gamma }Q_{\alpha \beta }\right)- \frac{4\pi}{p} L\epsilon_{\alpha \beta \gamma }Q_{\alpha \rho } \partial_{\gamma }Q_{\beta \rho }
\nonumber\\
&- \left[K\left(\delta(z)+\delta(z-N_z)\right)+E^2 \Delta \epsilon \right] Q_{zz}.
\label{FE}
\end{align}
The  first line is the Landau--de Gennes~\cite{deGennes,Grebel1983} thermal term describing the nematic to isotropic second order phase transition in temperature (parameters $a$, $b$, and $c$ are chosen to ensure a reasonable value for $S$, see Supp. Mat. (SM) ). In the second line,  elastic energies (of single elastic constant $L$)  penalize the gradient of $Q$ and favor a twist with cholesteric pitch $p$. The last line reflects the homeotropic surface anchoring of strength $K$ at the boundaries ($z=0, N_z$) and the coupling to a uniform electric field $E$, applied in the $z$ direction. $\Delta\epsilon$ is the dielectric anisotropy of the LC favoring easy-axis (along $z$) or easy-plane (perpendicular to $z$) alignment depending on its sign. We will express the coefficients of the alignment terms, $K$ and $\alpha=\Delta \epsilon E^2$, in dimensionless units by setting $\alpha_0=LSq_0^2=1$ and $K_0=LSq_0=1$ where $q_0$ is the natural twist. Then, $K/K_0=p/2\pi\xi_K$ and $\alpha/\alpha_0=(p/2\pi\xi_E)^2$ where $\xi_K=LS/K$ is the anchoring extrapolation length and $\xi_E=\sqrt{LS/\alpha}$ is the electrostatic coherence length. Expressing Eq.~(\ref{FE}) in dimensionless terms (see SM) reveals that the ratios $\alpha/\alpha_0$ and $(K/K_0)/(N_z/p)$ determine the alignment strength.

Frustration in the form of alignment in the vertical direction can be used to stabilize particle-like solutions called skyrmions. Fig.~\ref{fig1}a shows the mid-plane of one such {\it full skyrmion}, where the polar angle of the director $\vec{n}$ rotates by $180^\circ$ from its center to periphery leading to a topological charge $\frac{1}{4\pi}\int\int \mathrm{d}x \mathrm{d}y\, \vec{n}\cdot(\partial_x \vec{n} \times \partial_y \vec{n} ) =1$, as the mapping of the directors to the surface of a sphere covers the surface once. 
In Fig.~\ref{fig1}b, we show skyrmions with the same topology whose size and shape are controlled by $K$ and $\alpha$, for a  cell thickness $N_z=0.36p$. When $K\neq 0$, skyrmions form barrel-like spherulites which can be fully embedded inside the cell. 
As $K$ is further reduced and $\alpha$ becomes dominant, the $z$ dependence becomes small, as also seen in experiments~\cite{ackerman2014two,tai2019surface}. The special case of $K=0$ yields a $z$-invariant structure~\cite{duzgun2020alignment,DeMatteis2018}, and can be modeled as 2D allowing the possibility of simulating very large systems. Finally, Fig.~\ref{fig1}c demonstrates size control via the electric field which was previously studies in detail \cite{duzgun_SkPhase}. Main text includes only 2D simulations and full 3D simulations are presented in SM.

Crucially, unlike {\em merons}~\cite{Fukuda2011, Nych2017, duzgun_SkPhase, active_merons}, LC Skyrmions are \emph{local} objects (not accompanied by defects) that can be generated and decimated at will as long lived isolated particles--they neither disappear, nor spontaneously appear~\cite{Foster2019_SkBags,duzgun2020alignment,berteloot2020ring-shaped,lavrentovich2020design}. They can be actuated, and arranged to exhibit a variety of {\it collective} dynamics~\cite{Sohn19optics,ackerman2017squirming,Sohn2018_cargo}.
 Skyrmions can be confined, and manipulated~\cite{duzgun_SkPhase} via  light, electric fields and surface chemistry~\cite{ackerman2014two,channels_wei,toron_in_channel,electric_generation,tai2019surface}. 
They are attracted by regions of weak (and repelled by regions of strong) easy-axis alignment~\cite{duzgun2020alignment,duzgunSM2020}. Also, skyrmions are repelled from regions exposed to light which increases the helical pitch $p$~\cite{Sohn2017APS}. Finally, confinement by electrical field is not made problematic by fringe effects or lack of sharp gradients in real systems, as we have shown~\cite{duzgunSM2020,duzgun2020alignment}.

\begin{figure}[t!]
\includegraphics[width=0.85\columnwidth]{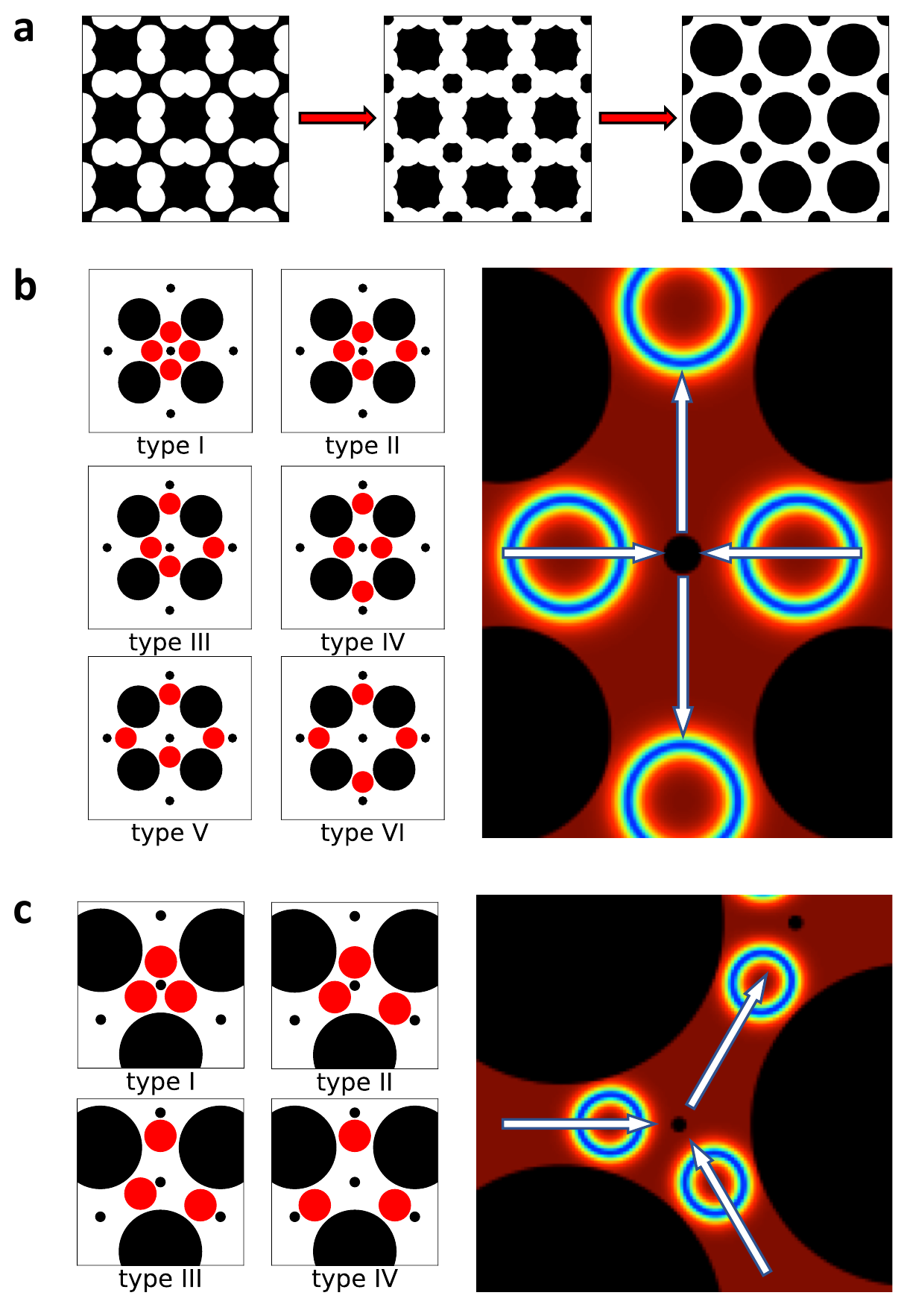}
\caption{{\bf Constructing Binary Variables out of Trapped Skyrmions}. \textbf{(a)} Dumbbell traps with closed ends would suppress skyrmion-skyrmion {interaction hence} are modified to final design with open ends. \textbf{(b)} Schematics of vertex configuration types in order of decreasing energy for square skyrmionic ice. Red circles represent the skyrmions, and vertex types are listed  in order of decreasing energy. Ice-rule configurations are the two Type IV vertices corresponding to 2 skyrmions in the vertex, and two out of the vertex. {\bf (c)} Vertex configurations in order of decreasing energy for hexagonal skyrmionic ice. Ice-rule configurations are the three Type II vertices (two skyrmions in, one out) and 3 Type III (one in, two out) vertices.} 
\label{fig2}
\end{figure}

\begin{figure*}[t!]
\includegraphics[width=.85\linewidth]{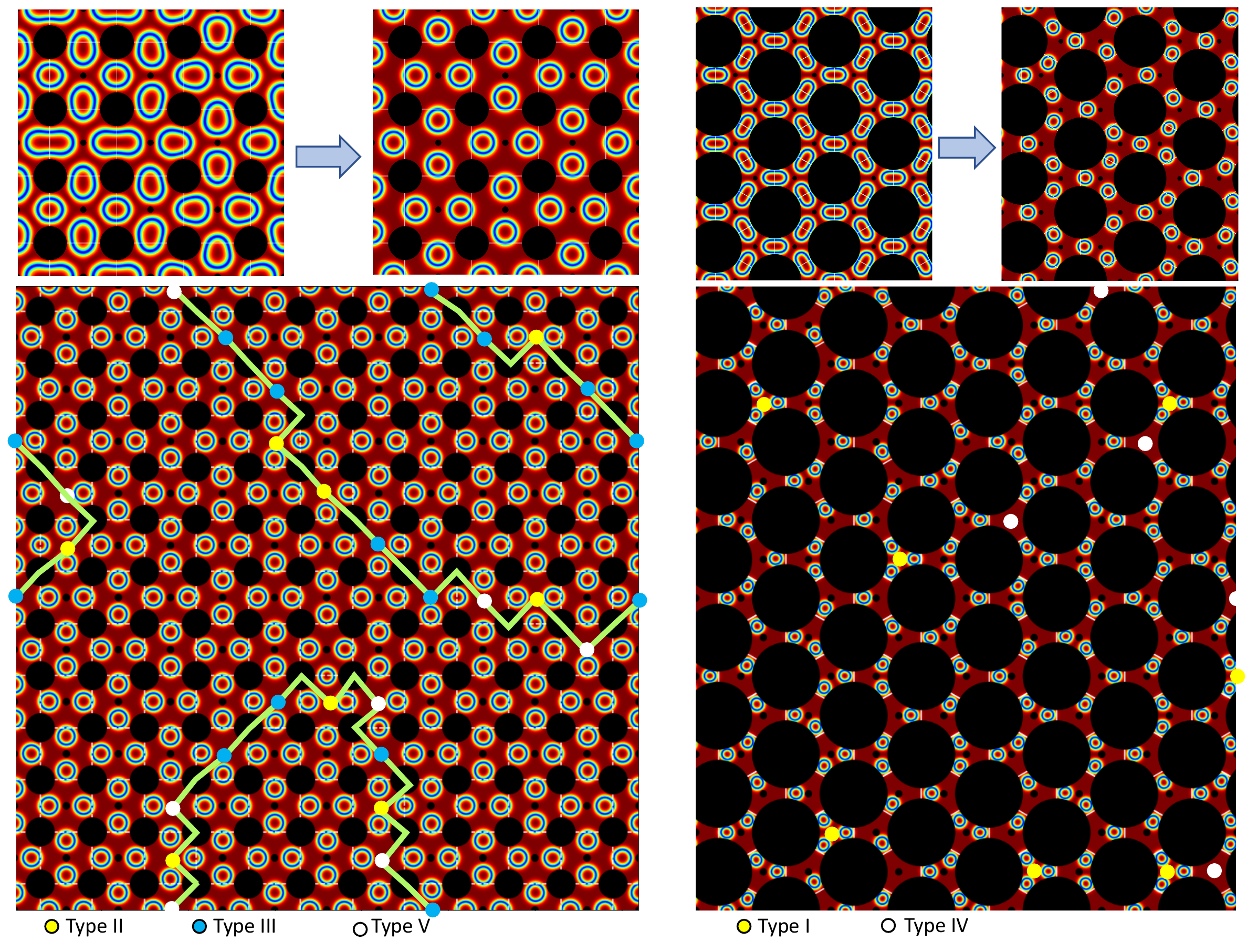}
\caption{{\bf Relaxed skyrmion spin ices obey the ice rule.} \textbf{Top:} Skyrmions initially swollen and then de-swollen, relax to lower energy states. \textbf{Bottom:} 2D square and hexagonal ice with surface anchoring/field relax to ice-rule obeying states with defects. The square geometry shows an ordered state where domains of different ``antiferromagnetic'' orientations of ice-rule, type IV vertices are separated by domain walls (green lines) of ice-rule obeying Type III (blue) and ice-rule violating Type II (yellow) and Type V (white). The hexagonal geometry converges to a disordered manifold of ice-rule obeying Type II (2-in/1-out) and  Type III (1-in/2-out) vertices. Excitations are ice-rule violating Type I  (3-in, yellow) or Type IV (3-out, white) vertices. White lines superimposed on the lattices  highlight the midpoint of the traps. 
Aspect ratios of the traps are 0.52 for square and 0.41 for hexagonal lattices. Other simulation parameters are listed in supplementary materials.}
\label{fig3}
\end{figure*}

To build a spin ice, we need to confine skyrmions in binary traps that can be considered pseudo-spins~\cite{ortiz2019colloquium}, using the mechanisms above. The task is non-trivial. Previous works on colloids~\cite{Libal2006,ortiz2016engineering} suggests  a dumbbell-shaped confinement (Fig.~2a). This choice {\it would not} work for skyrmions because traps with closed ends would suppress their mutual elastic interaction. The second panel of Fig.~\ref{fig2}-(a) shows how to go from dumbbells  to our much simpler and general geometry with {\em open ends}. There, smaller black circles represent trap ends and bigger circles provide the narrower mid-section of the trap. 

The usual nomenclature~\cite{ortiz2019colloquium} for skyrmions' configurations at the vertices are shown in Fig.~\ref{fig2}b-c along with their spin representation for a square and hexagonal lattice respectively. It is expected~\cite{Libal2006}, as a result of non-local frustration~\cite{nisoli2018unexpected}, that collective lowest energy states  obey the {\it ice rule}~\cite{bernal1933theory,Pauling1935}: 2 particles in the vertex and 2 out, for the square geometry and 1-in/2-out or 2-in/1-out for the hexagonal one. Further, the square geometry should lead to an ordered state and the hexagonal to a disordered one.

 \begin{figure*}[t]
\includegraphics[width=0.93\linewidth]{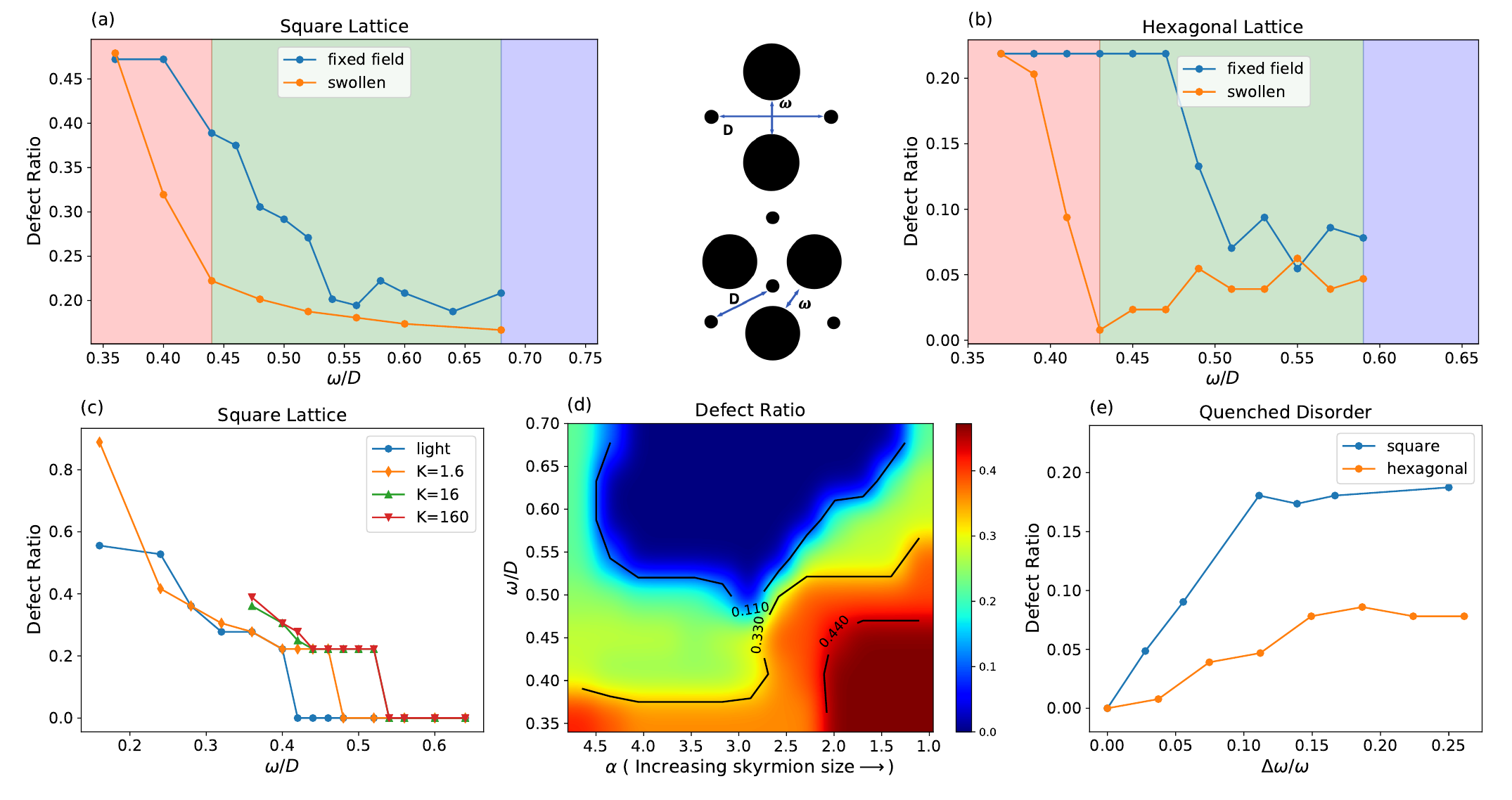}
\caption{ {\bf Limits of the ice rule validity.} \textbf{(a,b)} Defect ratio vs. size of the central circle in the trap  for square and hexagonal lattices.  Purple region: $\omega$ values too large for trap asymmetry. Pink region: $\omega$ values too small to allow skyrmion motion. Green region: ice behavior observed. \textbf{(c)} Suitable obstacle size depends on the type of wall mechanism and its strength. Results for obstacles produced by light and by various strengths of surface anchoring $K$ inside the obstacles are shown. \textbf{(d)} Skyrmion size also influences how traps work. Ice rule is not observed for very large or very small skyrmions. \textbf{(e)} Defect counts vs. random noise added to the width of traps. The noise is uniformly distributed between $\pm\Delta\omega$ with $D=60$ and $\omega/D=0.6$. Other simulation parameters are listed in SM.}
\label{fig4}
\end{figure*}

We run simulations of the LC systems of 2D skyrmions by solving the over-damped dynamic equation
$\partial_t Q(\mathbf{r},t)  = -\Gamma\, {\delta F}/{\delta Q(\mathbf{r},t) }$
(where $F= \int f(\mathbf{r})\,d^3r$ and $\Gamma$ is the mobility constant) using a finite differences method and with periodic boundary conditions~\cite{duzgun_SkPhase,duzgun2020alignment,duzgunSM2020}. Traps are realized via either extra field or surface anchoring.  We initiate the system in an unrelaxed state with 288 skyrmions in the square geometry and 192 in the hexagonal geometry placed randomly inside the traps. This entails about three million elements updated at each time step, which we implement by exploiting  the intrinsic parallelism of GPUs (see SM). Systems were updated for $10^7$ time steps for the square and $3\times10^7$ time steps for the hexagonal lattice. At the beginning, we reduce the background field to swell the skyrmions until they occupy almost their entire traps (Fig.~\ref{fig3}, top and SM) so as to bring them in close interaction, and then we deswell the skyrmions and let the system relax.

Figure~\ref{fig3} shows snapshots of the final states for the two geometries. Square ice converges to  an ordered ``antiferromagnetic''~\cite{Libal2006,Zhang2013,Porro2013} tessellation of ice-rule-obeying type-IV vertices, with two skyrmions close to, and two away from, each vertex. Deviations from type IV  correspond to ice-rule obeying Type III, but also to violations of the ice rule in the form of {\it monopoles}~\cite{Castelnovo2008}, or Type II and V. Together, these excitations form familiar domain walls (drawn according to the method in ref.~\cite{nisoli2020topological}) among the two possible orientations of ``antiferromagnetic" order. Hexagonal ice also converges properly to an ice state where, unlike square ice,  a disordered mixture of Type II and Type III obey  the (pseudo) ice rule (1-in/2-out and 2-in/1-out), together with sparse monopole defects (Type I, IV). 

Structural parameters control proximity to the ice rule. Consider the aspect ratio of traps  $\omega/D$,  where $\omega$ is the width of the middle of the trap and $D$ is the length of the edge of the lattice.
If $\omega/D$ is too small (pink region in Fig~\ref{fig4}-(a,b)), the skyrmions are frozen in the trap; if too large (violet region), the trap is no longer binary and the skyrmions can sit at the center. The transition to this second regime is interesting as it can lead to the realization of a still largely unexplored classical spin-1 ice model where spins can be $-1,0,1$, and will be explored in future work. At intermediate values (green region) we are closest to the ice-rule.

The distinction between the three regions is fuzzy, also because the system preserves memory of its preparation. Figures~\ref{fig4}-(a,b) show that cyclically swelling and deswelling the skyrmions helps the system find lower energy states and extends the suitable range for ice configurations after 2-3 cycles, and leads to states closer to the ice rule.  More cycles do not change the final state. Figure~\ref{fig4}-(c), shows how curves of ice-rule violations vs. aspect ratio for square ice varies depending on obstacle properties. Obstacles generated by weaker anchoring ($K=1.6$) are softer and allows increased mobility for the skyrmions thus helps the system reach the ice-state. Also, note that $K=16$ and $K=160$ have identical effects, indicating that the effect saturates rapidly in $K$. Light exposure was modeled by reducing $q_0\to q_0/1.2$ and produces an effect similar to changing $K\to1.2K$ as the relevant ratio is $K/LSq_0$.

Another relevant structural parameter is skyrmion size. In previous work~\cite{duzgun_SkPhase} we found that unconfined skyrmions  exist for $3<\alpha<5.5$ (smaller size at larger $\alpha$). The confining effect of traps, however, allows for skyrmions to exist even for $\alpha=0$. It becomes then interesting to study the combined effect of trap aspect ratio and skyrmion size.
Figure~\ref{fig4}-(d) shows a contour plot of the defect count vs. $\omega/D$ and $\alpha$ demonstrating different regimes, including a blue area where defects are less than 10\%. In agreement with Fig.~4-(a), already discussed, the $\omega/D > 0.5$ region correspond to minimal defects. There, we observe an ample blue region where the ice configuration is reached easily and not affected much by the value of $\alpha$. When skyrmions become too small ($\alpha >4.5$), however, they do not interact, and we see more defects. On the opposite side, if skyrmions are too big compared with trap width they cannot properly move within the trap, and defects also increase. The demarcating line among regimes is therefore approximately linear.

Finally, previous spin ices are relatively robust against weak quenched disorder~\cite{budrikis2012disorder,budrikis2011diversity,budrikis2012disorder2}. To test our proposal we introduce disorder in the positions of the bigger circular walls while keeping the trap ends unchanged, leading to a random shift in $\pm\Delta\omega$.  Figure~4-(e) shows that disorder affects the square geometry more than the hexagonal. That is expected, as square spin ice has an ordered ground state and its excitations are strongly correlated via the domain walls of Fig.~3. Instead, sparse monopole defects in hexagonal spin ice are weakly correlated.

One last note. The reader will have noticed that the statistics of Fig.~4 (c,d,e) can reach zero defects whereas Fig.~4 (a) does not (intentionally chosen to illustrate connected unhappy vertices). The reason is that the system employed in Fig.~4 (c,d) is four times smaller and thus can more easily reach low energy states. In (e), the system size is the same as in (a) but the simulation time is 4 times as long (parameters are shown in SM).
 
We have demonstrated numerically LCs as a new platform for spin ice physics on the two most common gemoetries. In future works, we will explore extension  to more complex geometries~\cite{Morrison2013,nisoli2017deliberate,lao2018classical,farhan2017nanoscale}. Unlike previous platforms, LC can allow dynamic change of structure, for instance for cycling between topologically equivalent geometries of different ice behavior, for memory effects. Unlike trapped colloids, the skyrmions can change size, can be created or destroyed optically on the fly, to explore decimation, ice-rule fragility~\cite{libal2018ice}, or doping~\cite{libal2015doped}. Their mutual interaction can be controlled, from anisotropic to isotropic, by changing the direction of the background field~\cite{duzgun2019APS,Sohn2019clusters}. A growing abundance of techniques for controlling the collective behavior of LC topological defects suggest employmen  for actuation in soft robotics, optical applications, and functional materials design. Our proposal to realize spin ice with LC skyrmions is a promising development in this direction which we believe will stimulate experimental efforts. 
 
We wish to thank Ivan Smalyukh and Hayley Sohn (University of Colorado at Boulder) for useful discussions on skyrmion confinement and manipulation and Michael Varga for discussions on CUDA programming. 
This work was carried out under the auspices of the US Department of Energy through the Los Alamos National Laboratory, operated  by Triad National Security, LLC (Contract No. 892333218NCA000001).

\bibliography{LCspinice,library2.bib}

\end{document}